\documentclass[sn-basic]{sn-jnl}

\usepackage{graphicx}%
\usepackage{multirow}%
\usepackage{amsmath,amssymb,amsfonts}%
\usepackage{amsthm}%
\usepackage{mathrsfs}%
\usepackage[title]{appendix}%
\usepackage{xcolor}%
\usepackage{textcomp}%
\usepackage{manyfoot}%
\usepackage{booktabs}%
\usepackage{algorithm}%
\usepackage{algorithmicx}%
\usepackage{algpseudocode}%
\usepackage{listings}%
\usepackage{blindtext}%
\usepackage{setspace}%
\usepackage{color,soul}

\theoremstyle{thmstyleone}%
\theoremstyle{thmstyletwo}%

\theoremstyle{thmstylethree}%

\begin{document}

\title[Search and study of young infrared stellar clusters]{Search and study of young infrared stellar clusters}


\author*[1]{\fnm{Naira} \sur{Azatyan}}\email{nayazatyan@bao.sci.am}


\affil*[1]{Byurakan Astrophysical Observatory, 0213 Aragatsotn Prov., Armenia}


\abstract{The main aim of this paper is to study both the Interstellar Medium (ISM) and the young stellar population in the three star-forming regions, namely IRAS\,05137+3919, 05168+3634, and 19110+1045. The study of the ISM includes determination of the hydrogen column density (N(H$_2$)) and dust temperature (T$_d$) in the regions using Modified blackbody fitting. The main parameters of identified and classified young stellar objects (YSOs) belonging to the regions were determined comparing with the radiation transfer models. We also constructed a colour-magnitude diagram to compare the parameters of the YSOs with the results of the radiative transfer models. The three stellar populations appear to have formed under different scenarios. In the cases of IRAS\,05137+3919 and IRAS\,05168+3634, the age spread is considerably wider, suggesting that the stellar population likely emerged from independent condensations. In contrast, the third region comprises a pair of ultra-compact HII regions (UCHIIs), G45.12+0.13 and G45.07+0.13, with a notably smaller age spread. This hints at the possibility that these clusters originated from a single triggering event.}




\maketitle

\section{Introduction}\label{sec1}

The process of star formation remains ongoing throughout various stages of evolution of our galaxy and other galaxies, even in the present era \cite{Ambartsumian47}. It stands as one of the most pivotal processes that yield observable outcomes within galaxies.  Giant molecular clouds are birthplaces of stellar population \citep{Soderblom10}. Stellar clusters that remain embedded within their original molecular clouds hold a special significance. They provide insights into distinguishing the characteristics of stellar clusters related to their birth conditions from those derived from subsequent evolution \citep{lada03}. Therefore, there exists a genetic connection between young stellar objects (YSOs) and surrounding Interstellar Medium (ISM). It necessitates an integrated approach to study star-forming regions which implies determination of the main properties of already formed young stellar clusters and the surrounding environment. Moreover, the integrated approach to study of embedded stellar clusters can provide information about the initial star formation scenarios. When star formation within clusters is triggered, one would expect a limited age spread among the stars within the cluster. Conversely, in protocluster condensations that initiate star formation independently, individual clumps are likely to exhibit a broader age distribution \citep[e.g. ][]{Preibisch12}.

\begin{figure}
\centering
\includegraphics[width=0.48\linewidth]{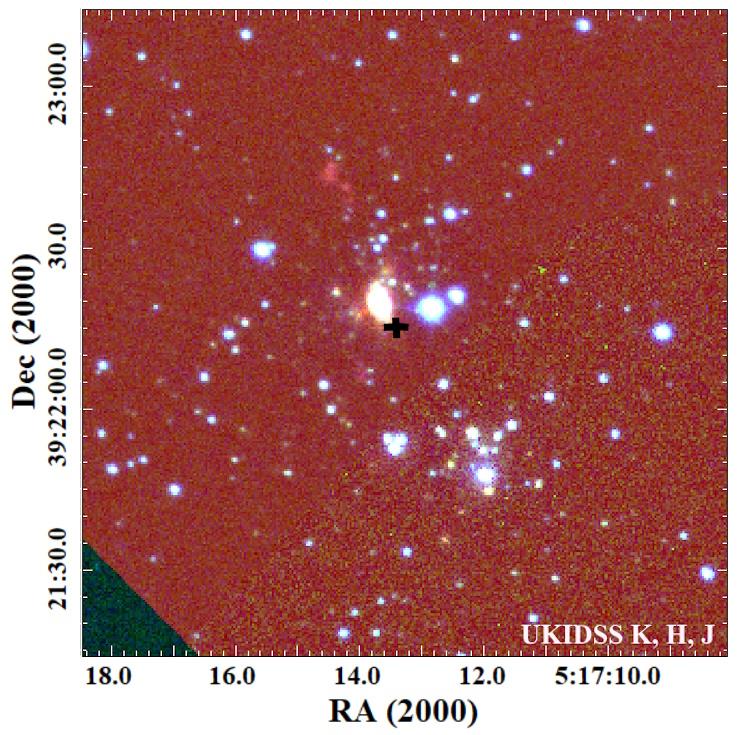}
\includegraphics[width=0.48\linewidth]{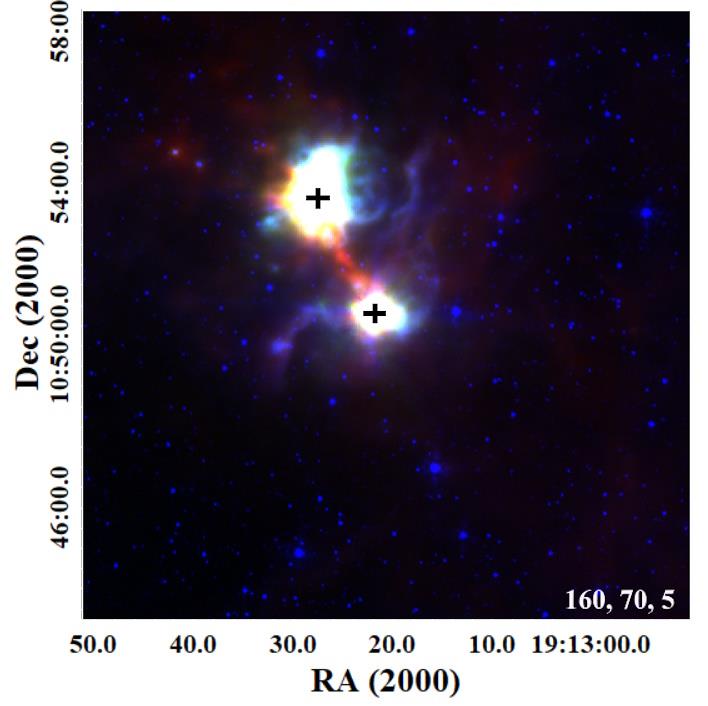}
\includegraphics[width=0.7\linewidth]{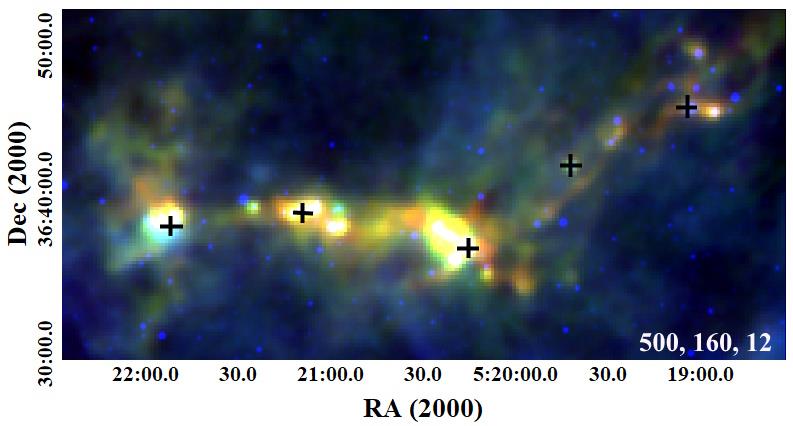}
\caption{(\textit{Left}) A colour-composite image of the IRAS\,05137+3919 star-forming region, combining data of UKIDSS J (in blue), H (in green), and K (in red). The position of the IRAS source is marked by a black cross. (\textit{Right}) A colour-composite image of the G45.12+0.13 and G45.07+0.13 UCHII regions, combining data of Spitzer 5.0\,$\mu$m (in blue), Herschel 70\,$\mu$m (in green), and Herschel 160\,$\mu$m (in red). The positions of the IRAS sources are marked by black crosses and they are in the following order from up to down: IRAS\,19111+1048 and 19110+1045. (\textit{Bottom}) A colour-composite image of the IRAS\,05168+3634 star-forming region, combining data of WISE 12\,$\mu$m (in blue), Herschel 160\,$\mu$m (in green), and Herschel 500\,$\mu$m (in red). The positions of the IRAS sources are marked by black crosses and they are in the following order from the left to the right: IRAS\,05184+3635, 05177+3636, 05168+3634, 05162+3639, and 05156+3643.}
\label{fig:regions}
\end{figure}

This paper presents findings of detailed investigation of young stellar clusters in the three star-forming regions: IRAS 05137+3919, 05168+3634, and 19110+1045 with a substantial size and multi-component structures. All three regions are active knots for star formation. Figure \ref{fig:regions} shows colour-composite images of the three star-forming regions with combinations of different wavelength range images. For each of these regions, an exhaustive analysis was undertaken, covering the following topics: 1) determining ISM parameters, specifically the distribution of hydrogen column density (N(H$_2$)) and dust temperature (T$_d$); 2) the search for young stellar clusters; 3) identifying cluster members by examining their infrared characteristics; and 4) determining the age and age spread among the members within these clusters. This paper is essentially a generalization of the results presented in a series of works devoted to the above-mentioned topics \citep{Nikoghosyan14,azatyan16,Azatyan18,Azatyan19,Nikoghosyan20,Azatyan20,Nikoghosyan21,Azatyan22} which were the basis for my PhD thesis.

We have organised the paper as follows. Section \ref{2} describes the used data and methods; in Section \ref{3}, we analyse the stellar population and ISM properties in the regions. Finally, the study results are summarised in Section \ref{4}.

\section{Used data and methods}
\label{2}

The study encompasses two primary scientific focuses: the examination of the ISM and the analysis of YSOs. For this purpose, we used data covering a wide range of near- to far-infrared (NIR, FIR) wavelengths.

\begin{figure}
\centering
\includegraphics[width=0.8\linewidth]{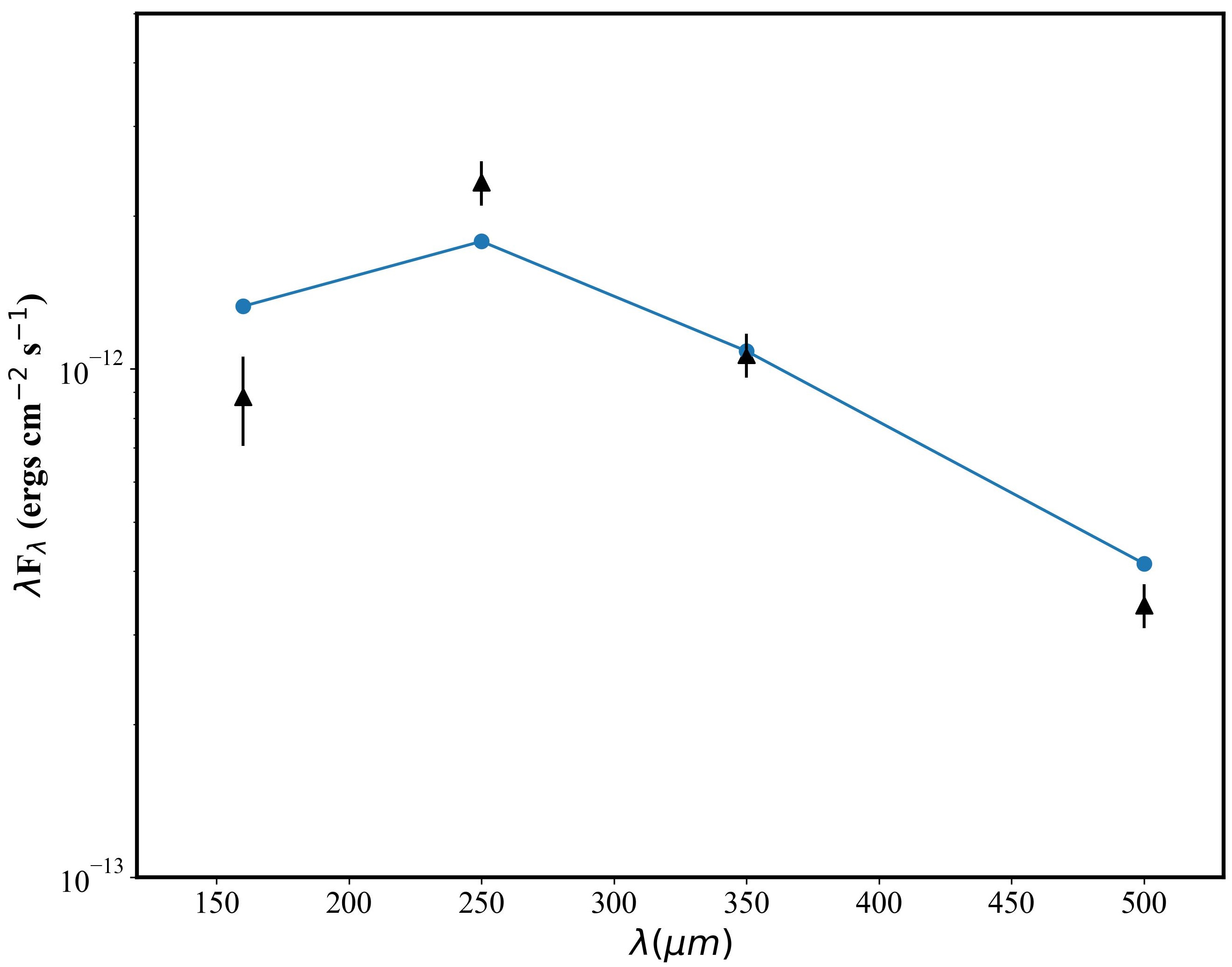}
\caption{A blackbody fitting of a low dust temperature (11\,K) in IRAS\,05137+3919 star-forming region. The filled triangles represent the observed fluxes in \textit{Herschel} four bands with their error bars, while filled circles represent the fluxes predicted by the model in the same bands. The blue line shows the fit of the predicted fluxes. The uncertainties between two samples is 0.024.}
\label{fig:SED}
\end{figure}

For the determination of N(H$_2$) and T$_d$ in the three star-forming regions, we used \textit{Herschel} FIR images, covering the optically thin spectral range of 160–500\,$\mu$m obtained on the Photodetector Array Camera and Spectrometer \citep[PACS,][]{poglitsch10} and the Spectral and Photometric Imaging Receiver \citep[SPIRE,][]{griffin10}. These spectral ranges are most productive for determining the parameters of the cold ISM in molecular clouds. Dust emission in the FIR can be modelled as a modified blackbody I$_{\nu}$=k$_{\nu}\mu_{H_2}m_{H}N(H_{2})B_{\nu}(T_{d})$, where $k_{\nu}$ is the dust opacity, $\mu_{H_2}$\,=\,2.8 representing the mean weight per hydrogen molecule \citep{Kauffmann08}, $m_{H}$ is the mass of hydrogen, $N(H_{2})$ is the hydrogen column density, and  $B_{\nu}(T_{d})$ is the Planck function at the dust temperature $T_{d}$. The dust opacity in the FIR is usually parametrized as a power law normalised to the value $k_{0}$ at a reference frequency $\nu_{0}$, so k$_{\nu}$\,=\,k$_{0}(\nu/\nu_{0})^{\beta}\,cm^{-2}g^{-1}$. Following \citet{hildebrand83}, \citet{Konyves15}, we adopted k$_{0}$\,=\,0.1\,cm$^{-2}g^{-1}$ at 300\,$\mu$m (gas-to-dust ratio of 100) and $\beta$\,=\,2. \citet{roy2014} showed that this assumption is good within better than 50\% in the bulk of molecular clouds. Based on the discussion in previous studies \citep[e.g.][]{Battersby11}, we excluded the 70\,$\mu$m data as the optically thin assumption may not hold. In addition, the emission here would have a significant contribution from a warm dust component, thus, modelling with a single-temperature blackbody would over-estimate the derived temperature. The three shortward intensity maps were degraded to the spatial resolution of the 500\,$\mu$m band; the four maps were then translated to a common coordinate system using 14\"\, pixels at all wavelengths. The SED fitting procedure for the cold ISM was executed pixel by pixel. Following to \citet{pezzuto2021}, we created a grid of models by varying only the temperature, in the range 5\,$\le$\,T$_{d}(K)$\,$\le$\,50, in step of 0.01\,K and for each temperature $T_{j}$, the code computes the intensity at FIR wavelengths for all bands. Since the intensity $I_{\nu}$ is linear with $N(H_{2})$, we can compute the column density at each pixel using a straightforward application of the least-squares technique. The uncertainty of $I_{\nu}$ for SPIRE is 10\% and for PACS is 20\% \citep{Konyves15}. Figure \ref{fig:SED} shows an example of blackbody fitting for low dust temperature (11\,K) in IRAS\,05137+3919 star-forming region. The uncertainties between the observed and predicted fluxes is 0.024 for this particular pixel.

For the second task that is the analysis of YSOs, we used data spanning the NIR, mid-infrared (MIR), and FIR wavelengths. For this task, it was used the data of Galactic Plane Survey DR6 \citep[UKIDSS GPS,][]{lucas08}, Two Micron Sky Survey \citep[2MASS,][]{Cutri03}, Galactic Legacy Infrared Midplane Survey Extraordinaire (GLIMPSE) and GLIMPSE\,360  \citep{churchwell09}, Multiband Infrared Photometer for {\itshape Spitzer} \citep[MIPSGAL,][]{Carey09}, Wide-field Infrared Survey Explorer \citep[WISE,][]{wright10}, Midcourse Space Experiment \citep[MSX, ][]{Price01}, {\itshape Herschel} PACS, SPIRE, {\itshape Herschel} Infrared Galactic Plane Survey \citep[Hi-GAL,][]{molinari16}, and the Infrared Astronomical Satellite \citep[IRAS,][]{Neugebauer84} Point Source Catalog v2.1 (PSC) catalogs. UKIDSS-DR6 catalog was selected as a primary. Then, this catalog was cross-matched with other MIR and FIR catalogs, using a matching radius of 3\,$\sigma$, accounting for the combined error. Since, the presence of circumstellar discs and envelopes cause an infrared (IR) excess of a YSO, therefore YSO candidates were identified based on their position in colour–colour (c-c) IR diagrams. The choice of colours relied on the availability of data. We employed two c-c diagrams, specifically (J-H) versus (H-K) and (J-K) versus [3.6]-[4.5], to identify YSOs within the IRAS\,05137+3919 star-forming region. Then, the selection of YSOs in the IRAS\,05168+3634 star-forming region was guided by the analysis of four c-c diagrams, specifically (J-H) versus (H-K), K-[3.6] versus [3.6]-[4.5], [3.4]-[4.6] versus [4.6]-[12], and [3.4]-[4.6] versus [4.6]-[22]. Finally, the selection of YSOs in the IRAS\,19110+1045 star-forming region involved the utilization of six c-c diagrams, specifically (J-H) versus (H-K), K-[3.6] versus [3.6]-[4.5], [3.6]-[4.5] versus [5.8]-[8.0], [3.6]-[5.8] versus [8.0]-[24], [3.4]-[4.6] versus [4.6]-[12], and [3.4]-[4.6] versus [4.6]-[22]. To validate the selected YSOs and extract their essential characteristics, we constructed their spectral energy distributions (SEDs) and compared them with radiative transfer models developed by \citet{Robitaille17}. Fig. 7 in \citet{Azatyan19} shows examples of such a comparison for the YSOs associated with IRAS sources in the IRAS\,05168+3634 region.

\begin{table}[h]
\caption{The parameters of ISM and stellar population in the star-forming regions}
{\fontsize{6.5}{7}\selectfont \label{tab:par}
\begin{tabular*}{\textwidth}{@{\extracolsep\fill}lcccccccc}
\toprule%
IRAS & $\alpha$\,(2000)  & $\delta$\,(2000)  & N(H$_2$) & T$_d$  & YSOs & Radius & D & A$_v$\\
 & (hh mm ss) & (dd mm ss)  & ($\times$\,10$^{23}$\,cm$^{-2}$)  & (K)  &  & (pc) & (kpc) & (mag)\\
 \midrule
(1) & (2) & (3) & (4) & (5) & (6) & (7) & (8) & (9)\\
\midrule\midrule
05137+3919  & 05 17 13.3 & +39 22 14.0 & 0.25\,$-$\,1.0 & 11\,$-$\,22 & 33 & 1.9/4.8 & 4.4/11 & 1.8\\
\midrule
05184+3635  & 05 21 53.2 & +36 38 20.4  & 1.1\,$-$\,1.5  & 12\,$-$\,15 & 52 & 1.4 & 1.9 & 1.4\\
05177+3636  & 05 21 09.4 & +36 39 37.1  & 1.1\,$-$\,2.3   & 12\,$-$\,13  & 79 & 1.9 & 1.9 & 1.3 \\
05168+3634  & 05 20 16.4 & +36 37 18.7  & 1.1\,$-$\,3.8   & 11\,$-$\,24 & 57 & 1.7 & 1.9 & 4.5\\
05162+3639  & 05 19 38.4 & +36 42 25.0 & 0.9\,$-$\,1.0   & 11\,$-$\,12  & 5 & 0.14 & 1.9 & 1.2\\
05156+3643  & 05 19 03.6 & +36 46 15.7 & 1.1\,$-$\,1.6   & 11\,$-$\,12  & 47 & 1.5 & 1.9 & 1.0\\
\midrule
19110+1045 & 19 13 22.0 & +10 50 54.0 & 3.0\,$-$\,5.0  & 18\,$-$\,42 & 37 & 1.8 & 7.8 & 13\\
19111+1048 & 19 13 27.8 & +10 53 36.7 & 3.0\,$-$\,5.5  & 18\,$-$\,35 & 87 & 2.7 & 7.8 & 13\\
\botrule
\end{tabular*}

\footnotetext{Note: (1)-names of (sub-)regions, (2),(3)- coordinates of the IRAS sources, (4)- ranges of hydrogen column density, (5)- ranges of dust temperature, (6)- numbers of YSOs within the selected radii, (7)-the radii of (sub-)regions, (8)- adopted distances of the regions, (9)- the median interstellar extinctions estimated with SED fitting tool for IRAS\,05137+3919 and 19110+1045\&19111+1048 star-forming regions, and according to the COBE/DIRBE and IRAS/ISSA maps \citep{Schlegel98} for IRAS\,05168+3634 star-forming region.}
}
\end{table}

\section{Results and Discussion}
\label{3}
The selected regions (IRAS\,05137+3919, 05168+3634, and 19110+1045), among other things, hold particular interest due to their considerable distances, providing an opportunity to evaluate the capabilities of the databases we have access to. For the IRAS\,05137+3919, the distance estimates based on radio observations have a large variation: from $\sim$\,4.4\,kpc \citep{Casoli86,Wouterloot89} to 10.8\,kpc \citep{Molinari96}. The distance estimations of IRAS\,05168+3634 star-forming region are also different: 6.08\,kpc \citep{Molinari96} and 1.88$^{+0.21}_{-0.17}$\,kpc \citep{Sakai12}. Given the significant discrepancy in the distance estimates for the IRAS\,05168+3634 star-forming region, we attempted to identify YSOs within the \textit{Gaia}\,EDR3 database. The results obtained from the \textit{Gaia}\,EDR3 data provide additional support to that this star-forming region is situated at a distance of $\sim$1.9\,kpc \citep{Nikoghosyan21}. And finally, the distance estimate for the IRAS\,19110+1045 star-forming region is 7.75$\pm$0.45\,kpc \citep{Wu19}. Table \ref{tab:par} contains the parameters of ISM and stellar population in the star-forming regions and below are the results of their study.

\singlespacing

In the vicinity of \textbf{IRAS\,05137+3919}, we observed a young stellar cluster closely associated with the CPM\,15 YSO \citep{Campbell89} (Figure \ref{fig:regions} top left). This region exhibits various signs of active star formation, including maser emissions, as well as the presence of CO and H$_2$ outflows \citep{Zhang05, Varricatt10}. The results of modified single-temperature blackbody fitting are presented in (4) and (5) Columns of Table \ref{tab:par}. The maxima for both N(H$_2$) and T$_d$ coincide with the position of the IRAS source. Within determined 1.5\'\, radius, we identify 33 YSOs using two c-c diagrams. We were able to derive the parameters of these YSOs using the SED fitting tool. These selected YSOs exhibit a non-uniform distribution within the star-forming region, forming two distinct subgroups. One of these subgroups is situated around CPM\,15, while the second group encompasses a notable number of middle-mass objects, surrounded by gas-dust nebulae \citep{Nikoghosyan14}.

\singlespacing

The next young stellar cluster is concentrated around the \textbf{IRAS\,05168+3634} source. The presence of diverse maser emissions and the existence of $^{13}$CO cores provide compelling evidence of its ongoing star-forming activity \citep{Zhang05, Fontani10,Guan08}. Notably, in the FIR wavelengths, the region exhibits a more intricate and complex structure compared to its appearance in the NIR, as illustrated in bottom panel of Figure \ref{fig:regions}.  Our investigation of the whole star-forming region within the molecular cloud unveiled the presence of four additional IRAS sources at the same distance, namely IRAS\,05184+3635, 05177+3636, 05162+3639, and 05156+3643, which are embedded into a $\sim$26\,pc long cloud complex \citep{Azatyan19}. The results of modified single-temperature blackbody fitting are presented in Table \ref{tab:par} for each IRAS sub-region. Notably, the relatively warmer gas-dusty material leads to dense condensations around the IRAS objects \citep{Nikoghosyan21}. An exception is IRAS\,05162+3639 sub-region where the N(H$_2$) map reveals a distinct absence of regions with relatively higher density. Additionally, no cluster of YSOs has been identified around this source; instead, only five stars are present. Within determined radii, we identify YSOs using four c-c diagrams and the results are presented in (6) Column of Table \ref{tab:par} for each IRAS sub-region. Comprehensive tables containing information on the selected YSOs, including their NIR, MIR, and FIR photometry, as well as parameters obtained through the SED fitting tool, can be accessed via the VizieR On-line Data Catalog\footnote{The full tables are available in the VizieR On-line Data Catalog: J/A+A/622/A38} \citep{Azatyan19}. 

\singlespacing

The final star-forming region is associated with the \textbf{IRAS\,19110+1045} and \textbf{19111+1048} sources, which are commonly referred to as the \textbf{G45.07+0.13} and \textbf{G45.12+0.13} UCHII regions, respectively. This system serves as an ideal laboratory for the exploration of the initial phases of massive star formation and their impact on their surrounding environments. Figure \ref{fig:regions} top right panel shows colour-composite image of the region and it also indicates that these UCHII regions are interconnected by a relatively cooler bridge, strongly suggesting that they form a physically linked system. The results of modified single-temperature blackbody fitting are presented in Table \ref{tab:par} for each IRAS sub-region. In the G45.07+0.13 region, the IRAS source is slightly offset from the density peak, while in the G45.12+0.13 region, the position of IRAS\,19111+1048 coincides with the maxima of both T$_d$ and N(H$_2$). Let us note that in this star-forming region both T$_d$ and N(H$_2$) are the highest. Within determined radii, we identify YSOs using six c-c diagrams and the results are presented in (6) Column of Table \ref{tab:par} for each IRAS sub-region. As in the previous cases, we included objects classified as YSOs in at least two c-c diagrams into our list. However, due to the presence of two saturated areas in the MIR band surrounding the IRAS sources (IRAS\,19110+1045 with a 25$''$ radius and IRAS\,19111+1048 with a 50$''$ radius), objects within those regions were classified as YSO candidates based solely on the NIR c-c diagram and we conducted a visual inspection of the YSO candidates (95 YSOs) in the two MIR-saturated regions, as these objects are of particular interest due to their proximity to the UCHIIs. The comprehensive tables containing information on the selected YSOs, including their NIR, MIR, and FIR photometry, as well as parameters derived from SED fitting tool, can be accessed in the VizieR On-line Data Catalog\footnote{The complete dataset is accessible in the VizieR On-line Data Catalog: J/other/PASA/39.24}. Among considered three star-forming regions, massive stars were only detected in this one. In the G45.07+0.13 region, we identified two massive stars, while in the G45.12+0.13 region - four, and one of them with a mass of 9.4±4.3\,M$_{\odot}$, a temperature of 23\,000±11\,000\,K, and an evolutionary age of (2.5±1.2)x10$^6$ years was successfully identified as the NIR counterpart of the IRAS\,19111+1048 source \citep{Azatyan22}.

\begin{figure}
\centering
\includegraphics[width=0.7\linewidth]{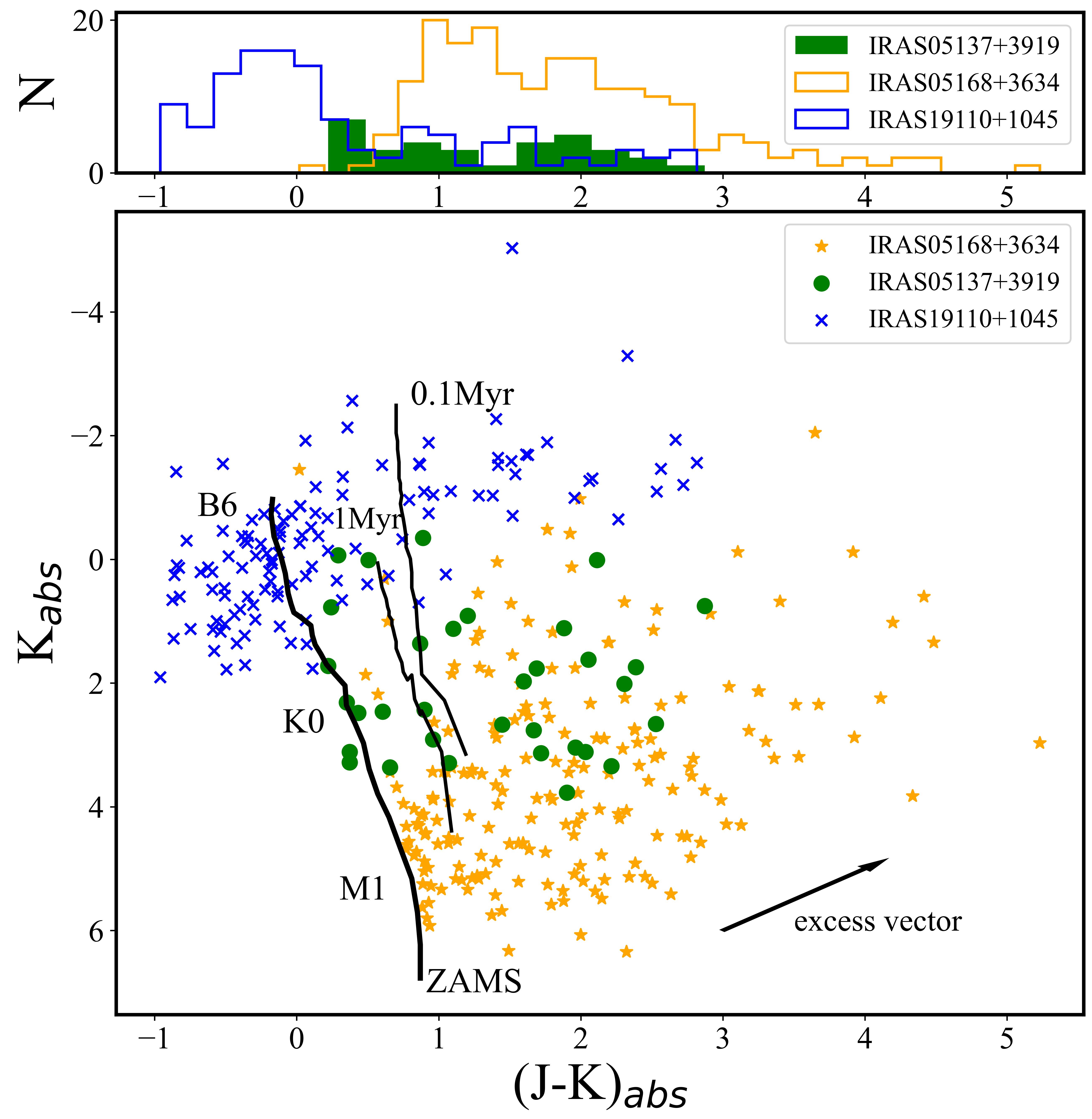}
\caption{(\textit{Bottom}) K versus (J–K) CMDs for the identified YSOs in the three star-forming regions. In these diagram, blue crosses represent stellar objects in the IRAS\,19110+1045 and 19111+1048 star-forming regions labelled as IRAS\,19110+1045, green circles represent objects located in the IRAS\,05137+3919 star-forming region, and finally yellow stars represent stellar objects in IRAS\,05168+3634 star-forming region. The pre-main sequence isochrones for the 0.1 and 1\,Myr \citep{siess00} and ZAMS are drawn as solid thin and thick lines, respectively. The positions of a few spectral types are labelled. The J and K magnitudes of the YSOs are corrected for the median interstellar extinctions and distances separately. The solid arrow indicates the average slope of NIR excesses caused by circumstellar discs \citep{lopez07}. (\textit{Top}) Histograms of (J - K)$_{abs}$ values of the three star-forming regions with the same colours.}
\label{fig:cmd19110}
\end{figure}

\begin{figure}
\centering
\includegraphics[width=0.7\linewidth]{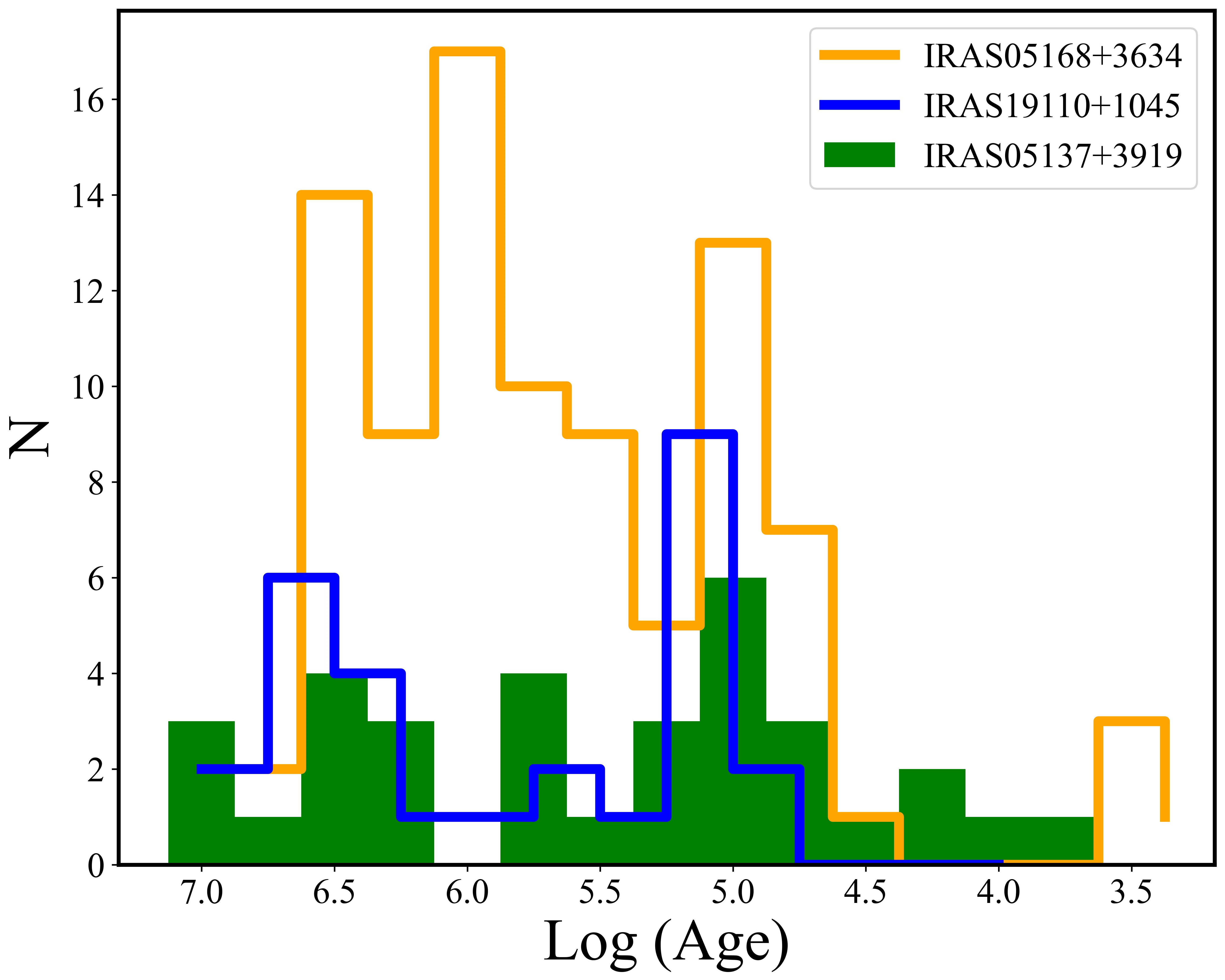}
\caption{Histograms of the evolutionary ages for members of the three star-forming regions. The colours are the same as in Figure \ref{fig:cmd19110}. The bin size corresponds to Log\,(Age) = 0.25.}
\label{fig:age}
\end{figure}

\singlespacing

Figure \ref{fig:cmd19110} bottom panel shows the distribution of the identified YSOs in the colour-magnitude diagram (CMD). The correction of J and K magnitudes of the YSOs for each region was done with the median interstellar extinctions and the adopted distances taken from Table \ref{tab:par}. For IRAS\,05137+3919, we used only 4.4\,kpc distance. The members of each star-forming regions are indicated with different colours and symbols. As we can see, the members of each star-forming regions display different dispersion concerning the isochrones (solid thin and thick lines). The members in IRAS\,05137+3919 and 05168+3634 star-forming regions show large dispersion concerning the isochrones, while the members in IRAS\,19110+1045 and 19111+1048 star-forming regions display relatively low dispersion concerning the isochrones and they are more evolved. To enhance clarity, the histograms for (J - K)$_{abs}$ are presented in the top panel with the same colours as in bottom panel. In general, the stellar objects in IRAS\,05137+3919 and 05168+3634 star-forming regions, display large spreads in (J - K)$_{abs}$, while the members in IRAS\,19110+1045 and 19111+1048 star-forming regions are mostly concentrated. For the comparison, we also constructed the histograms of the evolutionary ages for the members of the star-forming regions based on the results of the SED fitting tool. Figure \ref{fig:age} presents histograms of the evolutionary ages with the same colours as in Figure \ref{fig:cmd19110}. The histograms revealed a notably wide spread for all regions. It should be noted, the evolutionary age distribution for the objects in IRAS\,19110+1045 and 19111+1048 star-forming regions (blue bars) is derived from parameters obtained for only 29 YSOs subjected to the SED fitting tool. Most of the remaining 95 YSOs within the MIR-saturated regions are concentrated around the zero-age main sequence (ZAMS) and situated to the left of the 1\,Myr isochrone. Consequently, it is posited that these objects significantly contribute to the first peak in the evolutionary age distribution, rendering it a single, well-defined peak similar to the (J - K)$_{abs}$ histogram.  Consequently, the large age spread observed in the IRAS\,05137+3919 and 05168+3634 star-forming regions provides strong evidence to support the conclusion that the stellar populations within them are the outcome of independent condensations within the parent molecular cloud. In contrary, the small spread of evolutionary ages in IRAS\,19110+1045 and 19111+1048 star-forming regions suggests that formation of the stellar population may be attributed to a triggering shock. We assume that Granat 1915+105 High Mass X-ray Binary which is located at the same distance as IRAS\,19110+1045 and 19111+1048 star-forming regions and belongs to the same molecular cloud could be the possible trigger of the star formation in these regions. We plan to check the reliability of this assumption in our future work. 

\section{Conclusion}
\label{4}
Here are the main findings from the detailed study of the IRAS\,05137+3919, 05168+3634, 19110+1045\&19111+1048 star-forming regions:
\begin{itemize}
    \item A total of 33 YSOs were identified in the IRAS\,05137+3919 region, 240 YSOs in the IRAS\,05168+3634 region, and totally, 124 YSOs in the IRAS\,19110+1045 and 19111+1048 regions.
    \item In the IRAS\,05137+3919 star-forming region, the selected YSOs are not uniformly distributed and form two distinct subgroups.
    \item In the IRAS\,05168+3634 field, five dense subgroups were detected around IRAS sources, indicating that IRAS\,05168+3634 and four other sub-regions are embedded into a $\sim$26\,pc long cloud complex, located at $\sim$1.9\,kpc distance based on \textit{Gaia}\,EDR3 parallax measurements.
    \item Both IRAS\,05137+3919 and IRAS\,05168+3634 star-forming regions exhibit a wide age spread, suggesting that the stellar populations within these regions formed independently within the parent molecular clouds.
    \item The members of the IRAS clusters in the G45.12+0.13 and G45.07+0.13 UC\,HII regions show low scatter relative to the isochrones, and their age distribution indicates a small spread, possibly related to an external triggering shock.
    \item Among the considered star-forming regions, massive stars were only detected in the region where star formation may have been triggered and the temperature and hydrogen column density of the ISM are the highest, specifically IRAS\,19110+1045 and 19111+1048. 
    \end{itemize}
These findings provide valuable insights into the structure, formation, and age distribution of stars within these star-forming regions.

\backmatter

\bmhead{Acknowledgments}
\scriptsize{I thank the anonymous reviewer for constructive comments and suggestions. I would like to thank my supervisor – Dr. Elena Nikoghosyan for her invaluable supervision and support during the course of my PhD degree. This work partially was made possible by a research grant number №\,21AG-1C044 from Science Committee of Ministry of Education, Science, Culture and Sports RA.}

\bibliography{sn-bibliography}

\end{document}